\documentclass[a4paper]{article}

\usepackage{amssymb}
\usepackage{amsmath}
\usepackage{cases}
\usepackage{authblk}
\usepackage{geometry}
\title{Disk-sphere field duality theorem}

\author[1,2]{T. Maquart}
\providecommand{\keywords}[1]{\textbf{\textit{Keywords -}} #1}
\affil[1]{Universit{\'e} de Lyon, INSA-Lyon, CNRS UMR5259, LaMCoS, France}
\affil[2]{ANSYS Research \& Development, France}
\date{}

\affil[ ]{tristan.maquart@insa-lyon.fr}
\affil[ ]{tristan.maquart@ansys.com}

\begin{document}
\maketitle

\begin{abstract}
This paper presents a new reformulated theorem for fields embedded on a sphere or a disk. We focus in particular on the associated sphere of a disk when closing its only one boundary. We call this the disk-sphere duality theorem for the study of fields topological properties. For that purpose, we use the Poincar{\'e}-Hopf theorem and the boundary number theorem to firmly support our developments. In this context, the state of a $n$-symmetry direction field will be analyzed to show that disk-sphere field duality is closely related to the behavior near the disk's boundary.
\end{abstract}




\section{Introduction}
\label{sec1}

The study of fields on surfaces has a long interest beetween mechanical, physics and graphics communities. Topological properties of fields on surfaces are directly related to the $2$-dimensional manifold's topology. Applications exists in the graphics field for visualization purposes \cite{Kno13}. For many years, efforts have been spent to find a generalization of the Poincar{\'e}-Hopf theorem for higher dimensional manifolds \cite{Pug67,Fes92}.

In this paper, we present topological properties of fields embedded on $2$-dimensional manifolds in $\mathbb{R}^3$, especially for a disk along its only one boundary. We focus in particular on the sphere field properties \cite{Li06}. First, we introduce briefly topological concepts to securely anchor our further developments. Afterwards, we give mathematical field characteristics for a special case in a new reformulated theorem.

\section{Topology prerequisites}
\label{sec2}

In this section we provide prerequisites to understand the following developments. A theoretical background in surface topology is needed.

\subsection{Poincar{\'e}-Hopf theorem And Euler characteristic}
\label{sub1sec2}

The Poincar{\'e}-Hopf theorem explains the behavior of fields on compact differentiable manifolds. This theorem is widely used in geometry, physics, economics and other application fields. We give the Poincar{\'e}-Hopf theorem for a $2$-dimensional manifold $M$:\\

\textbf{Definition} : \textbf{Poincar{\'e}-Hopf theorem}. Let $M$ be a compact differentiable manifold and $\mathbf{d}$ be a 4-symmetry direction field with $n_s$ isolated singularities of indices $I_{\mathbf{d}}^i$ embedded in vertices $v$. If $M$ has some boundaries, the field must be pointing outward the normal direction along them:

\begin{equation}
\sum_{i=1}^{n_s} I_{\mathbf{d}}^i =\chi(M) .
\end{equation}

Where $\chi(M)$ is the Euler characteristic of the surface $M$. It is a topological invariant, an integer that describes the topological structure relative to the number of boundaries and the genus $g$ of the surface. We are interested only on compact oriented differentiable manifolds possibly with boundaries.

\subsection{Field singularities}
\label{sub2sec2}

The singularities of a vector field embedded on a surface are commonly a set of a finite dimension. If we define a vector field such as $\mathbf{d} : \mathbb{R}^2 \rightarrow \mathbb{R}^2$, the set of zeroes are the singularities, \textit{i.e.}, the set of $\mathbf{d}$ that respect : $\lbrace \mathbf{d}(x,y) = \mathbf{0} \rbrace$ for each entry. Depending of the symmetry of the field, the singularities can be classified by their index around a neighborhood $\Omega$ of points $P_i$ in place of singularities:

\begin{equation}
I_{\textbf{d}}(P_i)=\frac{1}{2 \pi} \int_{\partial \Omega (P_i)} d \theta .
\end{equation}

In addition, if we design a cycle $\gamma(s)$ to be equal to the boundary of the neighborhood $\Omega$, we can express the field singularities as:

\begin{equation}
I_{\textbf{d}}(P_i)=\frac{1}{2 \pi} \int_{\gamma (s) = \partial \Omega (P_i)} \kappa_{\textbf{d}} ds .
\end{equation}

Where $\kappa_{\textbf{d}}$ is the field curvature. The following development enable us to define correctly the indices of singularities on the boundaries $\partial M$ of a vector field $\mathbf{d}$. Thereafter, the next formulation is given in a generalized form using $n$-dimensional manifolds.\\ 

\textbf{Definition} : \textbf{Sum of singularity indices}. Let $\mathbf{d}$ be a vector field or a $n$-symmetry direction field with isolated zeros on the compact oriented differentiable $n$-dimensional manifold $M$, if $M$ has boundaries $\partial M$, the total index of singularity is defined to be the sum of its indices on the interior and on the boundaries \cite{Jub09}:

\begin{equation}
\sum_{i=1}^{n_s=n_i + n_b} I_{\textbf{d}}^i= \sum_{i=1}^{n_i} I_{\textbf{d}}^i + \sum_{i=1}^{n_b} I_{\textbf{d}}^i.
\label{eq:sum}
\end{equation}

Where $n_i$ represents the number of singularities embedded on surface whereas $n_b$ represents the number of singularities on boundaries.

\subsection{Field turning number}
\label{sub3sec2}

With the previous correct definitions of field singularities, we now describe the number of turns a field $\mathbf{d}$ can make along a given cycle $\gamma(s)$. It corresponds to the number of turns the field accomplish in a specific frame \cite{Ray08}. This amount of turns is called the turning number $T_{\textbf{d}}(\gamma)$ of $\mathbf{d}$ along the cycle $\gamma$.

\begin{equation}
T_{\textbf{d}}(\gamma)=\frac{1}{2 \pi} \int_{\gamma} (\kappa_{\textbf{d}} - \kappa_{\gamma}) ds .
\end{equation}

Where $\kappa_{\gamma}$ is the cycle geodesic curvature. We can reasonably show that the turning number in $\mathbb{R}^2$ can be also expressed with the index of singularity:

\begin{equation}
T_{\textbf{d}}(\partial \Omega (P_i))=I_{\textbf{d}}(P_i)-1 .
\label{eq:indexturn}
\end{equation}

\subsection{Field topological properties}
\label{sub4sec2}

Fields embedded on surfaces can contain relevant invariant information. Turning numbers have fundamental properties which make them useful to compare fields topologies. These information are straightforward to study fields on $2$-manifolds. Topology of a field is provided by turning numbers along boundary cycles, homology generators and around singularities \cite{Ray08}:\\

\textbf{Definition} : \textbf{Field topological equivalence}. Two direction fields defined over a surface $M$ are homotopic if and only if they have the same turning numbers along the cycles of their homology generators, boundaries and around singularities, yielding to the following statement:
 
\begin{equation}
\mathbf{d}_1 \equiv_t \mathbf{d}_2 \Leftrightarrow \forall \gamma \in H(M)=H_{g} (M) \cup \partial M, T_{\mathbf{d}_1} (\gamma )=T_{\mathbf{d}_2} (\gamma ) .
\end{equation}

Where $H_{g} (M)$ is the set of homology generators of $M$ whereas $\partial M$ is the set of boundary cycles. Singularities are omitted in this formulation. Notice that $\equiv_t$ denotes the topological equivalence.

\subsection{Boundary number theorem}
\label{sub5sec2}

Once we have determined turning numbers and topological properties of fields, we can now define the boundary number theorem. This theorem states the behavior of fields near boundaries depending to a topological invariant.\\

\textbf{Definition} : \textbf{Boundary number theorem}. Let $M$ be a compact differentiable $2$-manifold embedded in $\mathbb{R}^3$ with boundaries $\partial M$ and $\mathbf{d}$ be a $n$-symmetry direction field, then:

\begin{equation}
T_{\mathbf{d}} (\partial M) = -\chi(M) .
\end{equation}

Where $\chi(M)$ is the Euler characteristic of the surface $M$ and $\partial M$ is the set of boundaries. We can demonstrate that the boundary turning number theorem is equivalent to the Poincar{\'e}-Hopf theorem with a proper definition of the index of singularity \cite{Ray08}. For that purpose, we first generalize the index of singularity for a $2$-manifold embedded in $\mathbb{R}^3$ using cycle geodesic curvature $\kappa_{\gamma}$ and field curvature $\kappa_{\textbf{d}}$. We then formulate easily \eqref{eq:indexturn}:

\begin{equation}
I_{\textbf{d}}(P_i)=T_{\textbf{d}}(\partial \Omega (P_i))+1 .
\end{equation}

Afterwards, we store all the singularities in the associated closed mesh $M_c$ of $M$ in the new closed boundaries. This lead us to write the following equation involving $b$ boundary components:

\begin{equation}
T_{\textbf{d}}(\partial M)=T_{\textbf{d}}(\partial M_{c}) - \sum_i^b T_{\textbf{d}}^i (\partial \Omega (P_i))=- \sum_i^b (I_{\textbf{d}}^i (P_i) -1)=-\chi(M_c) +b .
\label{eq:boundaryturn}
\end{equation}

$T_{\textbf{d}}(\partial M_{c})=0$ due to the absence of boundaries.

\section{Disk-sphere field duality : opposite turning numbers}
\label{sec3}

This theorem is derived from the boundary turning number theorem and the Poincar{\'e}-Hopf theorem. A compact oriented differentiable $2$-dimensional manifold $D$ and $S$ are beeing considered. Let us introduce invariant integers for a topological disk surface $D$:

\begin{equation}
T_{\textbf{d}}(\partial D)=-\chi(D)=-\chi (D_c)+b=-1.
\end{equation}

With $T_{\textbf{d}}(\partial D)$ the turning number of a $n$-symmetry direction field $\textbf{d}$. $D_c$ is the associated closed mesh of $D$. In case of a topological sphere $S$, this turning number is equal to (consider an empty set $\dim(\partial S)=0$ of boundary components):

\begin{equation}
T_{\textbf{d}}(\partial S)=0 \neq -\chi(S)=-\chi(S_c)+b=-2 .
\end{equation}

Let us introduce all borders in place of singularities. Considering a neighborhood $\Omega$ around a point $P_i \in \partial D$ and also equations \eqref{eq:indexturn} and \eqref{eq:boundaryturn}, the turning number of the disk boundary can be formulated as:

\begin{equation}
T_{\textbf{d}}(\partial D)=-I_{\textbf{d}}(P_i)_{c}+1=-1 \Rightarrow I_{\textbf{d}}(P_i)_{c}=2.
\label{eq:sphere2}
\end{equation}

Where $I_{\textbf{d}}(P_i)_{c}$ is the index of singularity at $P_i$ on the closed surface $S$ of the associated disk $D$. Previous singularity index $I_{\textbf{d}}(P_i)_{c}$ can be decomposed into two different parts:

\begin{equation}
I_{\textbf{d}}(P_i)_{c}=\sum_{j=1}^2 I_{\textbf{d}}^j (P_{i})_{c}.
\end{equation}

One singularity is located on the boundary (\textit{i.e.}, on the associated sphere $S$), the other is embbeded in a disk vertex $v$ using sum of singularity indices in equation \eqref{eq:sum}:

\begin{equation}
I_{\textbf{d}}^1 (P_{i})_{c} \in S \text{ \& } I_{\textbf{d}}^2 (P_{i})_{c}=I_{\textbf{d}}(v) \in  D.
\end{equation}

Equation \eqref{eq:indexturn} is then evaluated for the boundary and also for the disk vertex in order to have the related turning numbers:

\begin{equation}
T_{\textbf{d}}(\partial \Omega (P_i))=I_{\textbf{d}}^1 (P_{i})_{c}-1 \text{ \& } T_{\textbf{d}}(\partial v)=I_{\textbf{d}}(v) -1.
\end{equation}

\textbf{Definition} : \textbf{Disk-sphere field duality theorem}. For two singularities on a sphere, one embedded in a vertex $v$ and one located at $P_i$, they have opposite turning numbers corresponding to the following duality:

\begin{equation}
T_{\textbf{d}}(\partial v)=I_{\textbf{d}}(v) -1=-(I_{\textbf{d}}^1 (P_{i})_{c}-1)=- T_{\textbf{d}}(\partial \Omega (P_i)).
\end{equation}

This is due to the definition of the sum of two singularity indices for a sphere $S$ in equation \eqref{eq:sphere2}. In other words, Poincar{\'e}-Hopf theorem violation on a disk is equivalent to define the right index of singularity on the associated sphere. It is possible to define a field without indices of singularity if at least one boundary exists. Therefore violations of Poincar{\'e}-Hopf theorem remains possible only if at least one boundary exists. This example states the trade between the boundary number theorem and the Poincar{\'e}-Hopf theorem. We use the term "violation" when not taking into account the behavior of the field near boundaries.

\section{Conclusion}
\label{sec4}

We have shown that for two singularities on a sphere, they have opposite turning numbers. This is a direct consequence of the Poincar{\'e}-Hopf theorem for topological spheres. The presented concepts are a re-formulation of the Poincar{\'e}-Hopf index formula, involving fields topological properties such as turning numbers. This was done in a specific case, when converting a disk into its associated topological sphere.

\appendix






\end{document}